\documentclass{aastex62}
\usepackage{color}
\usepackage{amsmath}
\usepackage{graphicx}
\usepackage{subfigure} 
\submitjournal{ApJ}

\begin{document}

\title{Model-independent determination of cosmic curvature based on Pad\'{e} approximation}


\author{Shi-Yu Li}
\affil{Department of Astronomy, Beijing Normal University, Beijing 100875, China}

\author{Yun-Long Li}
\affiliation{National Space Science Center, Chinese Academy of Sciences, Beijing 100190, China}

\author{Tong-Jie Zhang}
\affiliation{Department of Astronomy, Beijing Normal University, Beijing 100875, China; tjzhang@bnu.edu.cn}

\author{Tingting Zhang}
\affiliation{PLA Army Engineering University, Nanjing 210017, China; zhangtings@sohu.com}

\begin{abstract}
Given observations of the standard candles and the cosmic chronometers, we apply Pad\'{e} parameterization to the comoving distance and the Hubble paramter to find how stringent the constraint is set to the curvature parameter by the data. A weak informative prior is introduced in the modeling process to keep the inference away from the singularities. Bayesian evidence for different order of Pad\'{e} parameterizations is evaluated during the inference to select the most suitable parameterization in light of the data. The data we used prefer a parameterization form of comoving distance as $D_{01}(z)=\frac{a_0 z}{1+b_1 z}$ as well as a competitive form $D_{02}(z)=\frac{a_0 z}{1+b_1 z + b_2 z^2}$. Similar constraints on the spatial curvature parameter are established by those models and given the Hubble constant as a byproduct: $\Omega_k = 0.25^{+0.14}_{-0.13}$ (68\%  confidence level [C.L.]), $H_0 = 67.7 \pm 2.0$ km/s/Mpc (68\% C.L.) for $D_{01}$, and $\Omega_k = -0.01 \pm 0.13$ (68\% C.L.), $H_0 = 68.8 \pm 2.0$ km/s/Mpc (68\% C.L.) for $D_{02}$. The evidence of different models demonstrates the qualitative analysis of the Pad\'{e} parameterizations for the comoving distance.

\end{abstract}

\keywords{cosmology: cosmological parameters --- cosmology: observations}

\section{Introduction} \label{sec:intro}

The well-known geometry degeneracy makes it difficult to constrain curvature and dark energy simultaneously. One way to alleviate the problem is to assume the universe has a zero spatial curvature. This also helps simplify computational complexity. On the other hand, observations of cosmic microwave background (CMB) and baryon acoustic oscillation (BAO) impose strong constraints on the curvature in $\Lambda$CDM model, as a result extensive research regarding dark energy presuppose the curvature density is zero. However, \citet{Wright2007} reported that the assumption of a flat universe will lead to a constant dark energy Equation of State (EoS) vice versa. Reconstruction of dark energy EoS highly depends on the flatness of the universe. \citet{Clarkson2007} argued that even if there is little difference between the hypothetical cosmic curvature and the true value, there will be a large error at redshift $z \ge 0.9$. Therefore, the estimation of the spatial curvature is suggested to be carried out in a model-independent manner \citep{Bernstein2006, Clarkson2007, Oguri2012, Li2014, Rasanen2015, Cai2016, Yu2016, Li2016, Wei2017, Xia2017, Li2018, Denissenya2018}.

In fact, the curvature parameter $\Omega_k$ can be determined by the Hubble parameter $H(z)$, the comoving angular diameter distance $D(z)$, and the derivative of $D(z)$ at the same redshift without any assumptions of the dark energy EoS \citep{Clarkson2007}. The key work then turns to be the parameterization of the distance and the expansion rate. Taylor expansion has been widely used to approximate the luminosity distance \citep{Clarkson2010}. However, several literatures \citep{Aviles2014, Capozziello2017, Rezaei2017, Mehrabi2018} argued that Taylor polynomial expanding at $z = 0$ may diverge at high redshift thus they suggested using a rational polynomial. Pad\'{e} rational polynomial has the ability to fit any potential cosmological model with a better performance than Taylor expansion, due to its good convergency property in a relatively larger interval of redshift $z$. Therefore in this paper, we will use Pad\'{e} rational polynomial to obtain a continuous realization of $D(z)$ and its derivative to constrain the spatial curvature parameter.

The $D(z)$ modeling process using Pad\'{e} rational polynomials usually consists of two levels of inference: 1) model fitting - the inference of the coefficients of a Pad\'{e} expression with fixed orders, and 2) model selection - the inference of the order of the Pad\'{e} rational polynomial in light of the data. The second level inference which includes the evaluation of the model evidence is often computationally expensive. A possible way is to use the Akaike information criterion (AIC) and Bayesian information criterion (BIC) of model selection at present \citep{Kadane2004, Mehrabi2018}, since they are both approximations of the model evidence. But this method can introduce bias into the model selection. \citet{Kass1993} pointed out that BIC is biased towards simple models and AIC to complex models empirically. In the problem of using Pad\'{e} approximation to estimate curvature, which order belongs to the simple model or the complex model is not quantitatively described, thus the use of evidence is more suitable to find the best order of the Pad\'{e} approximation.

In next section, we use Pad\'{e} rational polynomial to build a family of parametric models. The scientific data and the likelihood of the data are described in Section \ref{sec:data}. In section \ref{sec:infer}, we calculate the inference of the coefficients and employ the Bayesian analysis to select the best model of  Pad\'{e} approximation for the estimation of the curvature parameter $\Omega_k$. We present the results in section \ref{sec:result} and provide the conclusions and discussions in section \ref{sec:conc}.

\section{The parametric model } \label{sec:method}
The relationship among the comoving distance $D(z)$, the Hubble parameter $H(z)$ and the curvature parameter $\Omega_k$ in the framework of the Friedmann-Robert-Walker metric is formulated as,

\begin{equation}
D(z)=\left\{
\begin{aligned}
&\frac{c}{H_0\sqrt{\Omega_k}} \sinh \left( \sqrt{\Omega_k} \chi(z) \right), & \Omega_k > 0 \\
&\frac{c}{H_0}\chi(z), & \Omega_k = 0 \\
&\frac{c}{H_0\sqrt{-\Omega_k}} \sin \left( \sqrt{-\Omega_k} \chi(z) \right), & \Omega_k < 0
\end{aligned}
\right.
\label{eq:D_vs_H}
\end{equation}
where $\chi(z) = \int_0^z{dz' \frac{H_0}{H(z')}}$, $H_0$ is the Hubble constant, $c$ is the speed of light. Solving for $\Omega_k$ from Eq.\ref{eq:D_vs_H} yields an explicit expression of the curvature parameter

\begin{equation}
 \Omega_k=\frac{[H(z)D'(z)]^2-c^2}{[H_0D(z)]^2},
 \label{eq:curv}
\end{equation}
which only requires the knowledge of the comoving distance and the Hubble parameter as well as the first derivative of the comoving distance with respect to redshift at the same redshift, thus providing a method to constrain the curvature parameter without any assumption of the dark energy EoS  \citep{Clarkson2007}.

Since the distance and the expansion rate can be derived by the observations of the standard candles \citep{Suzuki2012, Scolnic2018} and the cosmic chronometers \citep{Jimenez2002, Jimenez2003, Simon2005, Stern2010, Moresco2012, Zhang2014, Moresco2015, Moresco2016, Ratsimbazafy2017}, the key issue to apply Eq.\ref{eq:curv} to constrain the curvature is to find the estimation of $D'(z)$, which in this work is derived via the parameterization of the comoving distance using the Pad\'{e} rational polynomial.

\subsection{Pad\'{e} approximation} \label{subsec:pade}
The Pad\'{e} approximant of an arbitrary function $f(z)$ is given by the rational polynomial

\begin{equation}
 P_{mn}(z) = \frac{a_0 + a_1 z + \cdots a_m z^m}{b_0 + b_1 z + \cdots b_n z^n},
 \label{eq:pade}
\end{equation}
where the two non-negative integers, $m$ and $n$, are the degrees of the numerator and the denominator respectively. The coefficients $a_i (0\le i \le m)$ and $b_j (0 \le j \le n)$ are determined by solving the functions $P_{mn}^{(k)}(0)=f^{(k)}(0), (0 \le k \le m+n)$ if $f(z)$ has an explicit expression, or the coefficients can be derived by fitting $P_{mn}(z)$ to the data.

Let the comoving distance be approximated by the Pad\'{e} appoximant, $D(z) = c \cdot P_{mn}(z)$, where the constant $c$ is the speed of light. The condition $D(z=0)=0$ yields that $a_0 = 0$ and $b_0 \neq 0$, thus the numerator and the denominator of $P_{mn}(z)$ can be divided by $b_0$ and a factor $z$ can be extracted from the numerator. Rewrite the coefficients $\frac{a_{i+1}}{b_0}$ as $a_i (0 \le i \le m-1)$, $\frac{b_j}{b_0}$ as $b_j  (1 \le j \le n)$, and $m-1$ as $m$, then $D(z)$ can be parameterized as:

\begin{equation}
D(z) = c \cdot z \cdot
\frac{a_0 + a_1 z + \cdots a_m z^m}{1 + b_1 z + \cdots b_n z^n}.
\label{eq:pade_variant}
\end{equation}
Combining Eq.\ref{eq:curv} with Eq.\ref{eq:pade_variant} and the definition $H(z=0)=H_0$ yields that $H_0 = a_0^{-1}$, thus the Hubble parameter is parameterized as,
\begin{equation}
H(z) = \frac{\sqrt{\left[ a_0^{-1} D(z) \right]^2 \Omega_k + c^2}}{D'(z)}.
\label{eq:H}
\end{equation}

\subsection{Prior of the coefficients} \label{subsec:bayesian}
A Bayesian inference problem usually consists of its functional form, e.g. Eq.\ref{eq:pade_variant} and Eq.\ref{eq:H}, and the predictions the model makes about the data, e.g. the likelihood, as well as a prior distribution of the coefficients. It is quite common to apply a not very informative prior such as a wide flat prior to loosely bound the coefficients. However, due to the rational form, a randomly picked wide flat prior can not avoid the Pad\'{e} approximant to generate spurious singularities in the redshift range of the data, thus a weakly informative prior that regularize the smoothness of the Pad\'{e} approximant is needed to keep the inference in a reasonable range.

The singularities of the Pad\'{e} approximant (Eq.\ref{eq:pade_variant}) can be eliminated from the the range $z \in (0, \infty)$ by constraining the denominator to have no positive roots. This condition also states that the denominator is positive when $z > 0$. Given the constraint that the comoving distance $D(z)$ is positive in the range $z \in (0, \infty)$, the numerator must be positive in this range, leading to the requirement that the numerator has no positive roots either. The Descartes' rule of signs provides a simple sufficient but not necessary condition to construct a polynomial with no positive roots, that is to require all the coefficients of the polynomial to be non-negative. The smoothness constraint for the $H(z)$ is a little tough if using the Descartes' rule of signs, but if we consider the $H(z)$ as the output of a multi-layer perceptron whose weights are $\{a_i\} (0 \le i \le m), \{b_j\} (0 \le j \le n)$ and $\Omega_k$, a common prior that penalize the coefficients to achieve a smoother mapping can be proposed \citep{MacKay1992a, MacKay1992b},

\begin{equation}
-\log P(a, b, \Omega_k | m, n, \alpha, \mathcal{R}) = \alpha  \left(\sum_{i=0}^m \frac{1}{2}a_i^2 + \sum_{j=1}^n \frac{1}{2}b_j^2 + \frac{1}{2}\Omega_k^2 \right) + \log Z_w,
\label{eq:prior}
\end{equation}
where $\alpha > 0$ is the regularizing parameter of the simple quadratic prior $\mathcal{R}$, and $Z_w$ is the normalization constant which can be derived by integrating Eq.\ref{eq:prior} in the range  $a_i, b_j \in [0, \infty)$, $\Omega_k \in (-\infty, \infty)$,
\begin{equation}
\log Z_w = \frac{m+n+2}{2}\log\left(\frac{2\pi}{\alpha}\right) - (m+n+1)\log 2.
\end{equation}
If $\alpha$ approaches zero, Eq.\ref{eq:prior} returns to the flat prior. The best $\alpha$ is typically not known a priori, but later in section \ref{sec:infer} the value of $\alpha$ can be determined in light of the data via maximizing the evidence of the model.

\section{Data}\label{sec:data}
\subsection{The distances}
The comoving distance $D(z)$ is closely related to the distance modulus by $\mu(z) = 5 \log_{10} (1+z)D(z) + 25$. The latter is obtainable from the apparent magnitude $m = \mu + \mathcal{M}$ from the Pantheon supernovae samples \citep{Scolnic2018} which includes 1048 spectroscopically confirmed SNeIa. Here $\mathcal{M}$ is the absolute magnitude of a fiducial SNeIa. The likelihood of the dataset is defined as,

\begin{equation}
-\log\mathcal{L}_{SN}(m_{obs} \mid a, b, m, n) = \frac{\chi^2_{SN}}{2} + \log Z_{SN},
\label{eq:like_sn}
\end{equation}
where $\chi^2_{SN}$ is the modified misfit whose nuisance parameter $\mathcal{M}$ is already marginalized \citep{Conley2010},
\begin{equation}
\chi^2_{SN} = \boldsymbol{x}^T \left( \mathbf{\Sigma}_{SN}^{-1} - \frac{\mathbf{\Sigma}_{SN}^{-1} \mathbf{F}_1 \mathbf{\Sigma}_{SN}^{T^{-1}}}{\mathbf{1}^T \mathbf{\Sigma}_{SN}^{-1} \mathbf{1}} \right) \boldsymbol{x} + \log{\frac{\mathbf{1}^T \mathbf{\Sigma}_{SN}^{-1} \mathbf{1}}{2 \pi}}.
\label{eq:chi2_sn}
\end{equation}
Here $\mathbf{\Sigma}_{SN}$ is the covariance matrix of the apparent magnitude with systematics, $\mathbf{F_1}$ is an $1048 \times 1048$ matrix with each entry filled by 1. The vector $\boldsymbol{x}$ is defined as $\boldsymbol{x}= m_{obs} - \mu(z)$. $\mathbf{1}$ is a 1048-by-1 vector filled by 1. The gaussian integral gives the normalization constant $Z_{SN}$ as,
\begin{equation}
\log Z_{SN} = -\frac{1}{2}\log\frac{\mathbf{1}^T \mathbf{\Sigma}_{SN}^{-1} \mathbf{1}}{2\pi} + \frac{1048}{2}\log 2\pi -
\frac{1}{2}\log \det \left( \mathbf{\Sigma}_{SN}^{-1} - \frac{\mathbf{\Sigma}_{SN}^{-1} \mathbf{F}_1 \mathbf{\Sigma}_{SN}^{T^{-1}}}{\mathbf{1}^T \mathbf{\Sigma}_{SN}^{-1} \mathbf{1}} \right).
\label{eq:like_norm_sn}
\end{equation}

\subsection{The expansion rates}
Here 31 $H(z)$ data together with their errors are obtained from the tables in \citet{Cao2018}. These data are deduced from the cosmic chronometers in a cosmology model-independent approach described in \citet{Jimenez2002}. The BAO measurement is based on a fiducial cosmological model, thus $H(z)$ derived by BAOs are not included. All the Hubble parameter measurements are independent, the likelihood of the OHD has a simple form,

\begin{equation}
-\log\mathcal{L}_{H}(H_{obs} \mid a, b, \Omega_k, m, n) = \frac{\chi^2_H}{2} + \log Z_H
= \sum_{i=1}^{N_H}\frac{(H(z_i) - H_{obs, i})^2}{2\sigma_{H_i}^2} + \log Z_H,
\label{eq:like_h}
\end{equation}
where $N_H = 31$ and the normalization constant $Z_H$ is,
\begin{equation}
\log Z_H = \frac{N_H}{2}\log(2\pi) + \sum_{i=1}^{N_H}\log\sigma_i.
\end{equation}

\section{Inference of the coefficients and orders of the pad\'{e} approximant} \label{sec:infer}
The posterior distribution of the coefficients is simply given by Bayes' theorem,
\begin{equation}
P(w \mid D, m, n, \alpha, \mathcal{R}) =
\frac{\mathcal{L}_{SN}  \mathcal{L}_H \times P(w | m, n, \alpha, \mathcal{R})}{
P(D \mid m, n, \alpha, \mathcal{R})},
\label{eq:post}
\end{equation}
where $D$ stands for the data $m_{obs}$ and $H_{obs}$, $w$ stands for the coefficients $\{a_i\}$, $\{b_j\}$ and $\Omega_k$ in the model. $P(D \mid m, n, \alpha, \mathcal{R})$ is the normalization constant. The true posterior of the coefficients is defined by integrating Eq.\ref{eq:post}  over the regularizing parameter $\alpha$,
\begin{equation}
P(w \mid D, m, n, \mathcal{R}) = \int P(w \mid D, m, n, \alpha, \mathcal{R}) P(\alpha \mid D, m, n, \mathcal{R}) d\alpha.
\label{eq:post_true}
\end{equation}
The posterior $P(\alpha \mid D, m, n, \mathcal{R})$ usually has a strong peak at the most probable value $\hat{\alpha}$,  the integral above can be approximated by $P(w \mid D, m, n, \mathcal{R}) \approx P(w \mid D, m, n, \hat{\alpha}, \mathcal{R})$. The normalization constant in Eq.\ref{eq:post} is also the evidence how the data favors the model architecture $m, n$ with the regularization form $\mathcal{R}$ and its parameter $\alpha$. If there is no prior knowledge of $\alpha$, one can find the optimal $\hat{\alpha}$ by maximizing the value $P(D \mid m, n, \alpha, \mathcal{R})$.

By introducing a non-informative prior $P(\alpha)$ (since the $\alpha$ is a scaling factor, the prior is flat over $\log \alpha$) and integrating over $\alpha$, the final evidence of the model is obtained,
\begin{equation}
P(D \mid m, n, \mathcal{R}) = \int P(D \mid m, n, \alpha, \mathcal{R}) \times P(\alpha) d\alpha.
\label{eq:evidence_model}
\end{equation}
The evidence of the model determines which order $(m, n)$ of the Pad\'{e} approximant is the most probable in light of the data. It is the most difficult integral in this work but can be derived by the Laplace method whose key idea is to expand the integrant around the maximum posterior and approximate the integral by Gaussian integral \citep{MacKay2003, Kolokoltsov2018},
\begin{equation}
\ln P(D \mid m, n, \mathcal{R}) \approx \ln P(D \mid m, n, \mathcal{R}, \hat{\alpha}) + \ln P(\log\hat{\alpha}) + \frac{1}{2}\ln{2\pi} - \frac{1}{2}\ln \mathbf{A},
\label{eq:final_evidence}
\end{equation}
where $\mathbf{A} = -\frac{d^2}{d\alpha^2}\ln P(D \mid m, n, \hat{\alpha}, \mathcal{R})$. The error bound of the Laplace method is given by the Theorem 2 in \citet{Kolokoltsov2018}.

Using the prior proposed in Section \ref{subsec:bayesian}, $\mathbf{A}$ can be evaluated by sampling coefficients from the posterior $P(w \mid D, m, n, \hat{\alpha}, \mathcal{R})$,
\begin{equation}
\mathbf{A} = \left( \frac{m+n+2}{2} \right)^2 - \mathbb{E}_{P(w \mid D, m, n, \hat{\alpha}, \mathcal{R})}\left[ s^2 - s \right],
\end{equation}
where $s = \alpha  \left(\sum_{i=0}^m \frac{1}{2}a_i^2 + \sum_{j=1}^n \frac{1}{2}b_j^2 + \frac{1}{2}\Omega_k^2 \right)$. This work can be done by the nested sampling method \citep{Feroz2009} which has the ability to evaluate $P(D \mid m, n, \hat{\alpha}, \mathcal{R})$ and sample $P(w \mid D, m, n, \hat{\alpha})$ at the same time.

\section{Results} \label{sec:result}
\begin{figure}[htbp]
\centering
	\subfigure[]{
		\begin{minipage}[t]{0.5\linewidth}
		\centering
		\includegraphics[scale=0.4]{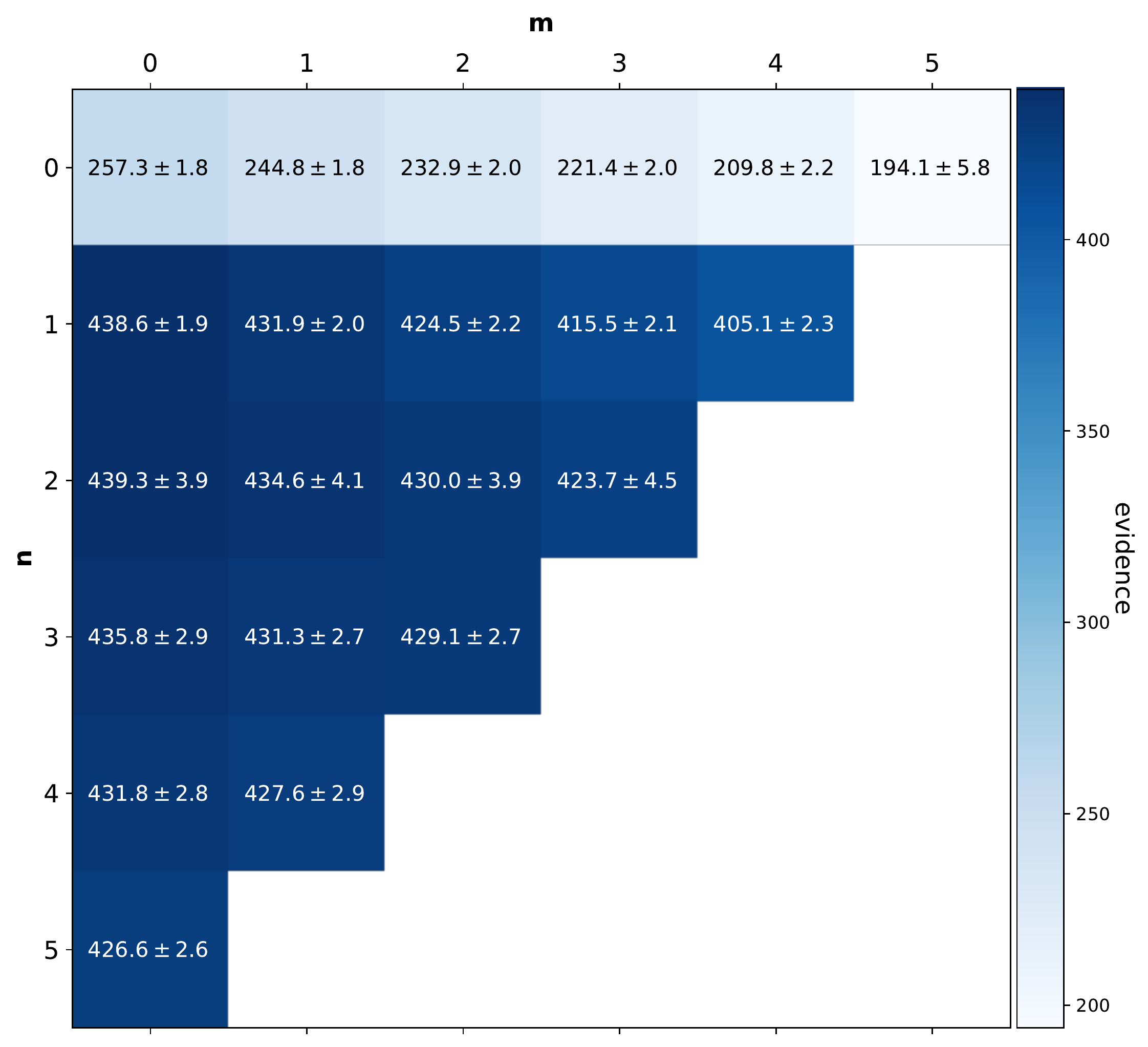}
		\end{minipage}%
		}%
	\subfigure[]{
		\begin{minipage}[t]{0.5\linewidth}
		\centering
		\includegraphics[scale=0.4]{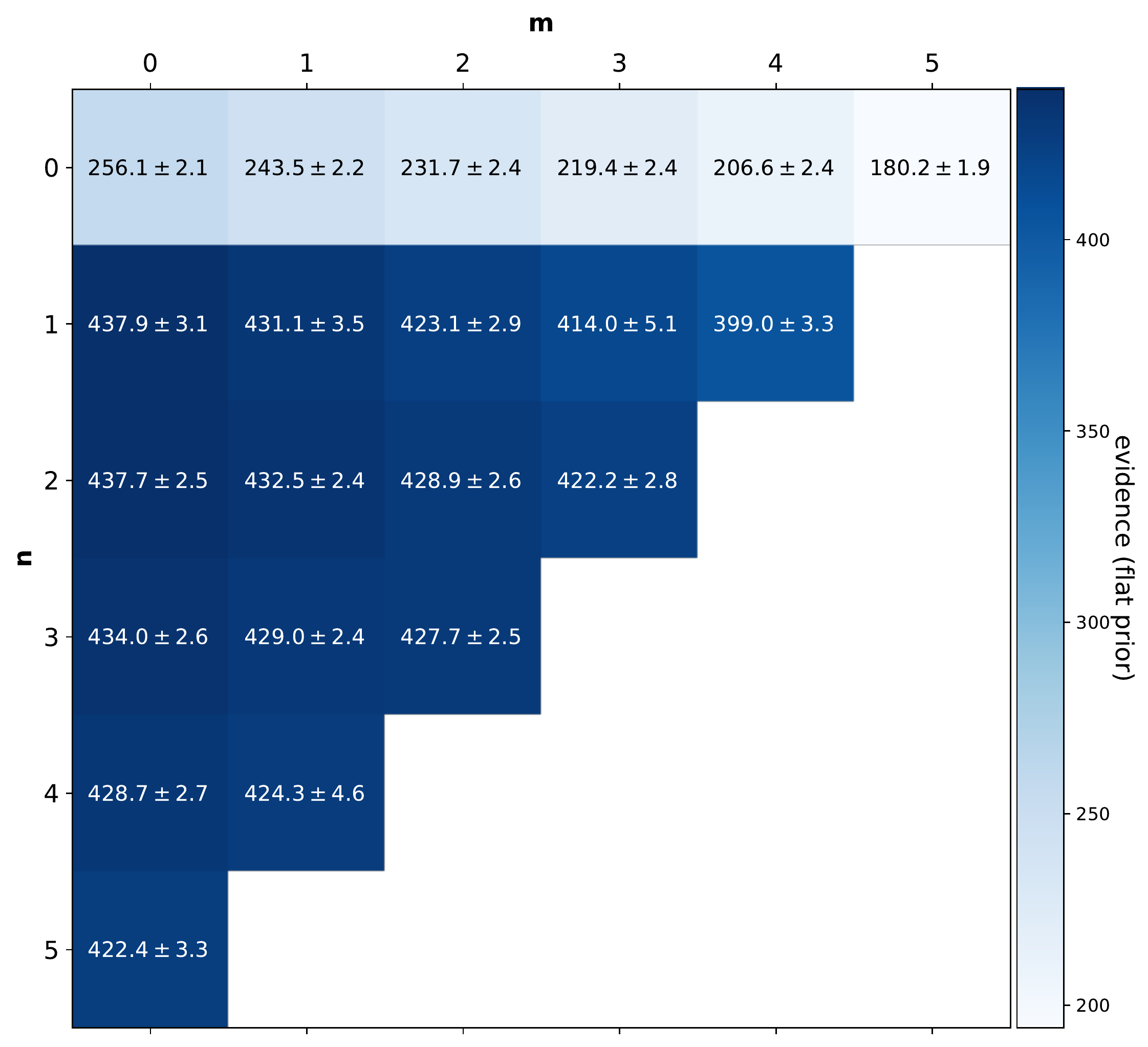}
		\end{minipage}%
}%
\centering
\caption{Log evidence of different Pad\'{e} approximants of order $(m, n)$. (a) evidence obtained using the proposed prior in Section \ref{subsec:bayesian}; (b) evidence obtained using the flat prior. Pad\'{e} approximants reduce to Taylor expansions when $n = 0$.}

\label{fig:evidence}
\end{figure}

By introducing a flat prior , $P(\log\alpha) = \frac{1}{12}\ (-2 \le \log\alpha \le 10)$, we have obtained the log evidence $E_{mn}=\log P(D \mid m ,n , \mathcal{R})$ of the parametric models built from the Pad\'{e} rational polynomial of the orders $(m, n), 0 \le m+n \le 5$, see Fig.\ref{fig:evidence} (a). A higher evidence indicates that the corresponding model is more preferred by the data. Notice that the Pad\'{e} approximants with $n=0$ actually reduce to the Taylor polynomials (see Eq.\ref{eq:pade_variant}), thus the first row in Fig.\ref{fig:evidence} shows the evidence of the model built from the Taylor polynomials. The difference between the model preferences, $\Delta E = E_{m'n'} - E_{mn}$, can be interpreted by the Jeffrey's scale, which was restated in \citet{Mehrabi2018}, to indicate how strong the evidence is against the model of order $(m, n)$ compared to the model of order $(m', n')$:  $\Delta E \in (0, 1.1)$ suggests weak evidence, and $\Delta E \in (1.1, 3)$ indicates definite evidence, while $\Delta E > 3$ means strong evidence.

The Pad\'{e} approximants show systematically better performance than the Taylor expansions with strong evidence. This result is consistent with the analysis from the perspective of convergence radius - Pad\'{e} approximation usually gives a better approximation than the corresponding truncated Taylor series over a large interval \citep{Aviles2014}. The model built from the Pad\'{e} approximant of order $(0,2)$ shows the highest evidence while the one of order $(0, 1)$ shows consistency with it under the Jeffery's scale. 

It is well known that the Bayesian evidence depends on the prior, a manually picked prior may introduce bias into model comparison, e.g., the selection of the orders $(m, n)$. However this source of bias in model comparison can be removed if prior covariances, e.g., $\alpha$, are estimated from data \citep{Penny2007}. As a comparison to the prior proposed in Section \ref{subsec:bayesian}, a flat prior $\mathcal{R}_f$ with a parameter $\alpha_f$ is applied in a parallel inference process,
\begin{equation}
P(w \mid m, n, \alpha_f, R_f)=
\left\{ 
\begin{aligned}
&\frac{1}{2\alpha_f^{m+n+2}}, &\text{if } 0 \le a, b \le \alpha_f,\text{and } \vert\Omega_k\vert \le \alpha_f \\
&0, &\text{otherwise}.
\end{aligned}
\right.
\end{equation}

The evidence of the orders using the flat prior are listed in Fig.\ref{fig:evidence}(b). The results are consistent to those in Fig.\ref{fig:evidence}(a), the bias in the selection of orders is removed. The Pad\'{e} approximants of order $(0, 1)$ and $(0, 2)$ still have the top two evidence. Thus we use these two Pad\'{e} approximant to find the curvature parameter respectively.

\begin{figure}[ht!]
\plottwo{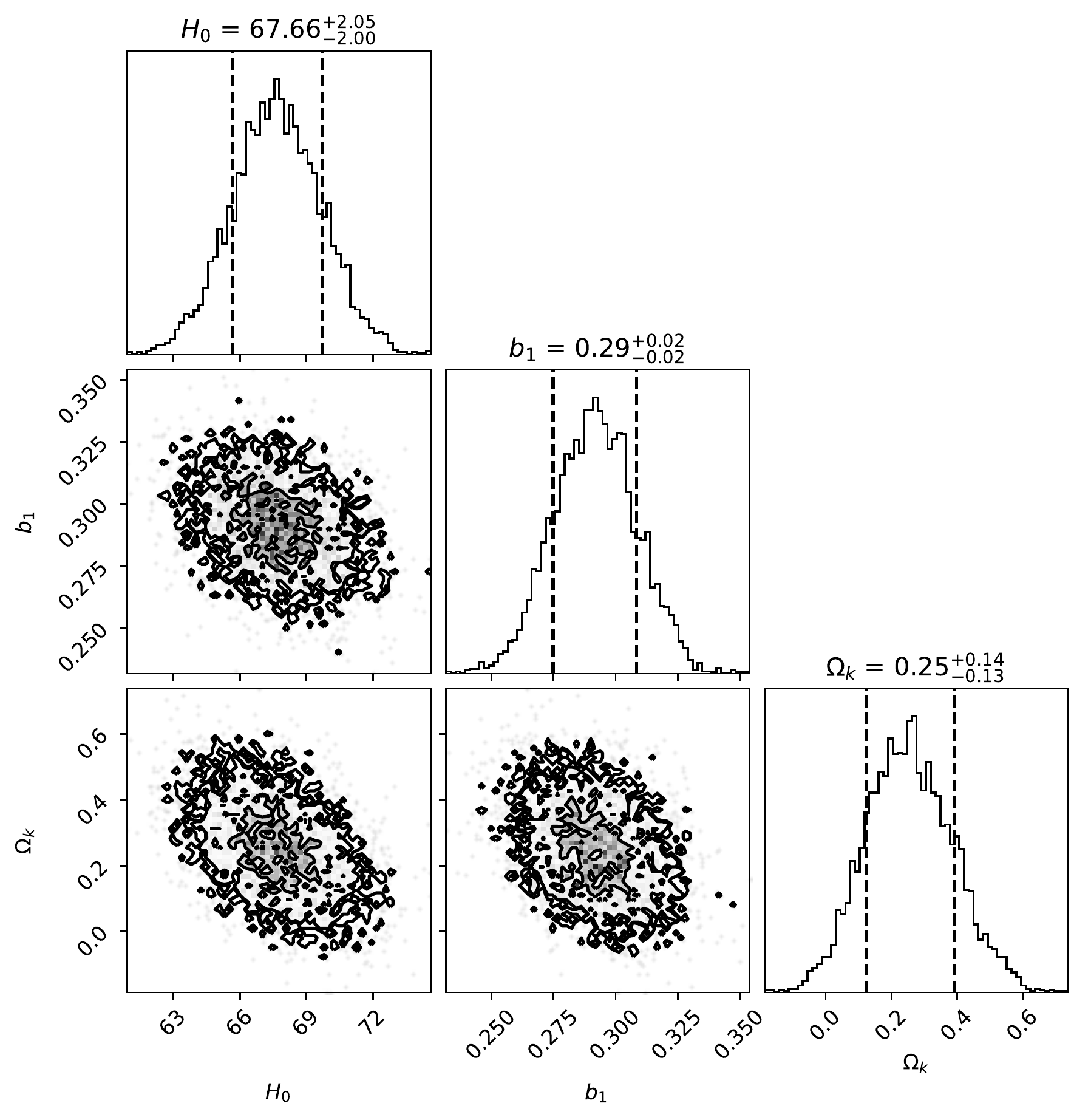}{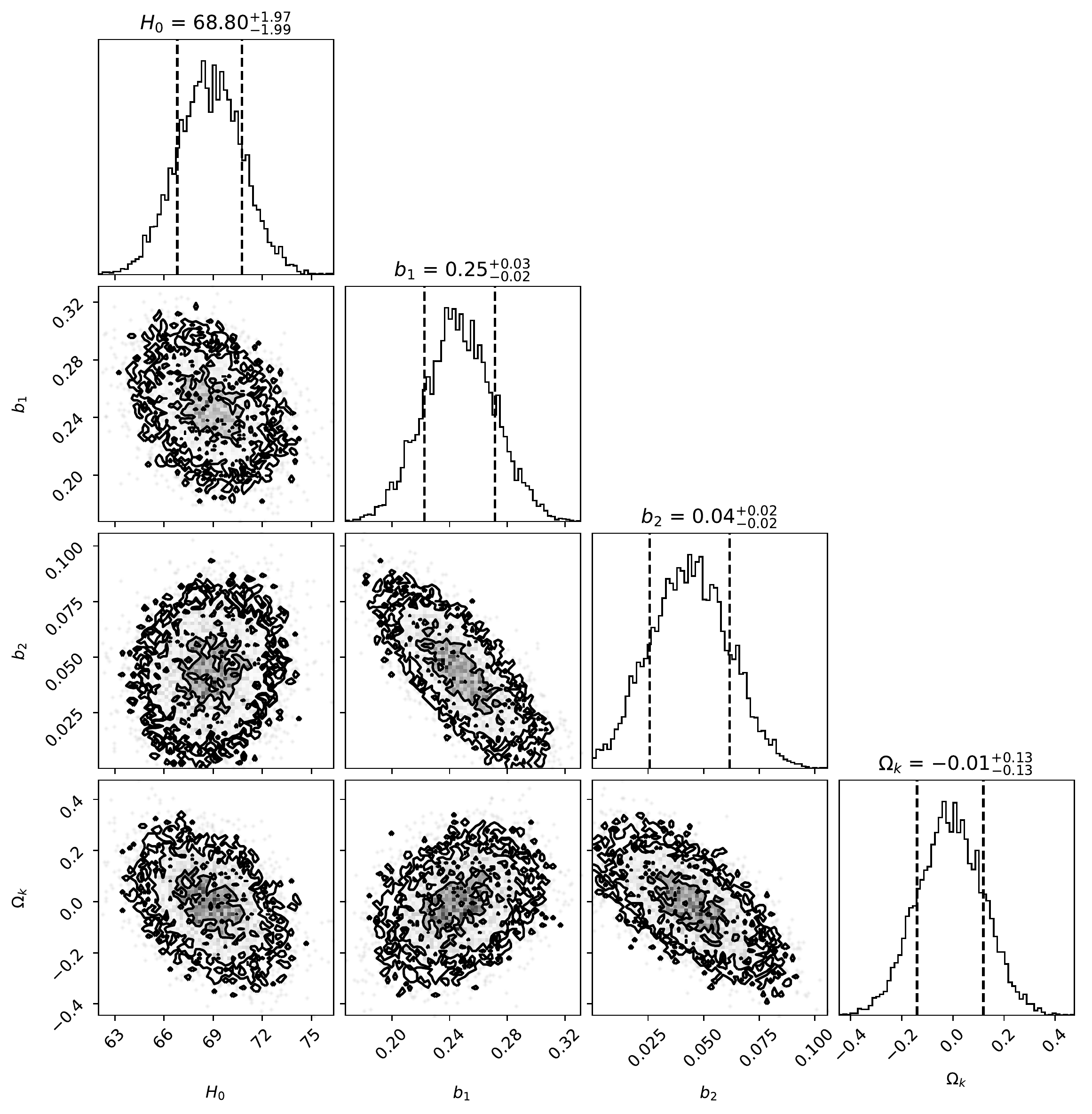}
\caption{The marginalized posterior constraints for the coefficients of the model built from the Pad\'{e} approximant of the order: (left panel) $m=0, n=1$; and  (right panel) $m=0, n=2$. The dashed lines mark the $1\sigma$ confidence level and the contour levels correspond to the $1\sigma, 2\sigma, 3\sigma$ confidence levels.}
\label{fig:parameter}
\end{figure}

The posterior in Eq. \ref{eq:post} is sampled by the \textit{pyMultiNest} package \citep{Buchner2014}, and the results are shown in Fig.\ref{fig:parameter}.
Both models constrain the curvature parameter with similar strength: the model of order $(0, 1)$ gives $\Omega_k = 0.25^{+0.14}_{-0.13}$ (68\% C.L.) and the model of order $(0, 2)$ gives $\Omega_k = -0.01 \pm 0.13$ (68\% C.L.). Although the model of order $(0, 1)$ prefers an open universe, it can not reject a flat universe with higher confidence. The flat prior with the best hyperparameter $\alpha_f$ gives the similar results, $\Omega_k = 0.24^{+0.08}_{-0.11}$ (68\% C.L.) for order $(0, 1)$ and $\Omega_k = -0.03^{+0.17}_{-0.15}$  (68\% C.L.) for order $(0, 2)$, once again indicating the prior covariances should be estimated from data.

Notice that the $\Omega_k$ in the posterior Eq.\ref{eq:post} is introduced by the likelihood of OHD and the simple quadratic prior. The latter which acts as a bound of coefficients with an effective size of $\alpha^{-\frac{1}{2}}$ contributes little to the inference of $\Omega_k$ compared with the likelihood of OHD, thus more precise OHD are expected to improve the constraint of $\Omega_k$ in this model-independent manner.
Since $H_0 = a_0^{-1}$, the constraint for the Hubble constant can be derived together with the curvature parameter as a byproduct: the model of order $(0, 1)$ gives $H_0 = 67.7 \pm 2.0$ km/s/Mpc (68\% C.L.) and the model of order $(0, 2)$ gives $H_0 = 68.8 \pm 2.0$ km/s/Mpc (68\% C.L.), both of which are in good accordance with the results from different analysis that employ the Pad\'{e} parameterization \citep{Rezaei2017, Mehrabi2018, Capozziello2019}.

\section{Conclusions and Discussions}\label{sec:conc}
In this paper, we build the parametric model of the comoving distance using the Pad\'{e} approximant (Eq.\ref{eq:pade_variant}), and derive the parametric model of the Hubble parameter with an additional parameter $\Omega_k$. During the modeling process, the Descartes' rule of signs is considered to exclude the singularities of $D(z)$ out of the range $z\in(0, \infty)$, that is to constrain the coefficients of Pad\'{e} polynomial to be nonnegative. However, since the parametric model of Hubble parameter involves the derivative of $D(z)$ and square root operation, applying this rule to the coefficients of $H(z)$ will lead to a parameter space too complex to evaluate the evidence of the model. Therefore a weakly informative prior of quadratic form with a free hyper parameter is applied to keep the smoothness of the model, and the hyper parameter is determined by the data adaptively. In this situation, the order of the parameterized model with the largest evidence is consistent with that of the best models discussed qualitatively in \citet{Aviles2014}. We could consider introducing such weakly informative prior to similar parametric model to explore a wider parameter space. Notice that this weakly informative prior does not exclude the possibility that parametric model of $H(z)$ produce singularities in the range $z\in(0, \infty)$. Singularities may even appear in the redshift interval spanned by the data, but the probability of occurrence is smaller than that of using an arbitrarily selected wide flat prior. Although Eq.\ref{eq:prior} is not a flat prior, it actually functions as a bound to the coefficients of the parametric model with an effective size $\alpha^{-\frac{1}{2}}$. The best regularizing constant $\hat{\alpha}$ is found adaptively in the range $\alpha\in(e^{-2}, e^{10})$ by maximizing the Bayesian evidence of this hyper parameter $P(D \vert m, n, \alpha, \mathcal{R})$. A typical value of $\hat{\alpha}$ is 17.7 for Pad\'{e} approximant and 0.73 for Taylor polynomial, and the corresponding equivalent widths are 0.2 and 1.2 respectively. It is consistent with the flat prior used in other Pad\'{e} parameterization to constrain the  cosmological parameters \citep{Mehrabi2018}. 

\citet{MacKay1992a} pointed out cubic spline prior, $-\log P(coef) = \alpha \int_{z_1}^{z_2}f''(z \mid coef)^2 dz$, might be more appropriate. It is exactly a quadratic prior in a linear model. Although the prior form becomes too complex to find a general analytical form in the nonlinear Pad\'{e} approximation model, especially when the order of the denominator of the Pad\'{e} approximant is larger than 2, the prior can completely avoid the singularity of the model in range ($z_1$,$z_2$). If we require the parametric model of $H(z)$ to have no singularity in the region of $z > 0$, it is necessary to apply the Descartes' rule of signs to establish a strong prior constraint, or employ Sturm theorem to establish a weaker but sufficient and necessary constraint, so that the numerator polynomial of $D'(z)$ in Eq.\ref{eq:H} has no positive roots. If we consider an expanding universe while $z > 0$, then the denominator polynomial of $D'(z)$ has no positive roots either, thus $D'(z)$ increase monotonously in this interval. Therefore, the order of Pad\'{e} rational polynomial must meet the condition that $n \le m+1$, in the denominator, coefficients of the items whose order exceed $m+1$ have to be $0$. For a given $m$, when $n$ gradually increases from $0$ to $m+1$, the evidence increases and reaches maximum, then starts to decrease. In fact, all the parametric models of $n > m+1$ are imitating the behavior of the parametric model of $n= m+1$, but the extra coefficients make the models penalized by the Occam's razor. This result shown in Fig.\ref{fig:evidence} is consistent with the analysis of the expanding universe. It is demonstrated once again that a weakly informative prior we adopt is suitable for the current problem.

The evaluation of the Bayesian evidence of the whole model suggests the most suitable parameterization of the comoving distance and the Hubble parameter should be constructed from the Pad\'{e} approximant of order $(0, 2)$, and a competitive model built from the Pad\'{e} approximant of order $(0, 1)$ is also noticed. With these two parametric models, the curvature parameter is constrained directly by the observations of the standard candles and the cosmic chronometers, and the Hubble constant is also constrained as a byproduct. Although the accuracy of the result is relatively low compared with other data analysis (such as BAO, CMB), it is still worthwhile to develop this method for more accurate data of Type Ia SNe and OHD \citep{Ma2011}. Since the evidence is the transportable quantity in Bayesian model comparisons, it can be applied to other parametric models such as principal component analysis (PCA) and non-parametric models such as Gaussian process, and learn how stringent constraint can be set to the curvature parameter by the data using the model with the best evidence.

\acknowledgments
This work was supported by National Key R\&D Program of China (2017YFA0402600), the National Science Foundation of China (Grants No. 11573006, 11929301, 61802428) and the 13th Five-year Informatization Plan of Chinese Academy of Sciences, Grant No. XXH13505-04.

\end{document}